\documentclass[aps,onecolumn,superscriptaddress,notitlepage,longbibliography]{revtex4-1}  
\usepackage{graphicx}  
\usepackage{epstopdf}
\usepackage{dcolumn}   
\usepackage{bm}        
\usepackage{amssymb}   
\usepackage{color}
\usepackage{ulem}

\newcommand{\Lnd}{\tilde{L}}
\newcommand{\Dtnd}{\tilde{D}_t}
\newcommand{\Drnd}{\tilde{D}_r}
\newcommand{\hpsi}{\hat{\psi}}
\newcommand{\hbpsi}{\hat{\bm{\psi}}}
\newcommand{\expe}{\mathrm{e}}
\newcommand{\mathi}{\mathrm{i}}

\usepackage{refstyle}
\usepackage{bm}
\usepackage{amsmath}
\usepackage{esint}
\begin{document}

\title{Stochastic Kinetic Theory for Collective Behavior of Hydrodynamically Interacting Active Particles }
\author{Yuzhou Qian}
\affiliation{The Howard P. Isermann Department of Chemical and Biological Engineering, Rensselaer Polytechnic Institute, Troy, New York 12180, USA}
\author{Peter R. Kramer}
\affiliation{Mathematical Sciences Department, Rensselaer Polytechnic Institute,
Troy, New York 12180, USA}
\author{Patrick T. Underhill}
\affiliation{The Howard P. Isermann Department of Chemical and Biological Engineering, Rensselaer Polytechnic Institute, Troy, New York 12180, USA}
\begin{abstract}
Self-propelled particles with hydrodynamic interactions (microswimmers) have previously been shown to produce long-range ordering phenomena. Many theoretical explanations for these collective phenomena  are connected to instabilities in the hydrodynamic or kinetic equations.  By incorporating stochastic fluxes into the mean field kinetic equation, we quantify the dynamics of a suspension of microswimmers in the parameter regime where the deterministic equation is stable.  We can thereby compute nontrivial collective phenomena concerning spatial correlations of orientation and stress as well as the enhanced diffusion of tracer particles.  Our analysis here focuses primarily on two-dimensional systems, though we also show how superdiffusion of tracers in three dimensions can occur by our framework.
\end{abstract}

\maketitle

\section{Introduction}
Bacteria, algal cells, and synthetic microscale motors can be conceptualized as hydrodynamically interacting self-propelled particles, which we will refer to as microswimmers \cite{Bergbook,BergBacteriaSwim1973,yilin_wu_et_al_self-organization_2011,koch_collective_2011,Pedley2010,RamaswamyReview2010,PowersReview2009,StockerReview2011,GoldsteinARFM2015}. The flows created by them can influence their own motion and the motion of passive objects suspended within the fluid. If the long-ranged hydrodynamics largely determine the behavior of such a system, the response can be understood by examining a suspension of dipole microswimmers.  In this article, we develop a fluctuation theory for a  model consisting of an idealized system of point dipole microswimmers.

Continuum field theories have been a fruitful approach to understanding collective phenomena in active suspensions with hydrodynamic interactions\cite{SubramanianRotationDiff2015,marchetti_hydrodynamics_2013,CateQuestion2012,CateLatticeMeanfield2011,GompperReview2015}. Such theories indicate that the linear stability of the uniform isotropic state depends on the sign of the force dipole, which classifies the ``puller'' and ``pusher'' mechanisms of swimming. The uniform isotropic configuration of pullers is always linearly stable, while for pushers it can suffer an instability when rotational diffusion is weak enough relative to the hydrodynamic coupling of orientations\cite{Saintillan2008,SaintillanDilute2009,Saintillanbook2012,Cateslatticeboltzmann2010,PEDLEYrevisit2010,Staintillan2011,ShelleyJRSI2012,Marchetti2010,Ramaswamy2002}.
Much attention has been focused on patterning phenomena proceeding nonlinearly from the instability, including the associated long-ranged correlations \cite{Staintillan2011,LaugaDense2011}, giant fluctuations \cite{Cate2DNonlinear2011,SwinneyDenseExp2010}, turbulence\cite{LowenDense2012} and other factors that influence the instability and nonlinear oscillatory dynamics \cite{Bozorgi2014,Bozorgi20142}. Indeed, the continuum mean-field equations describe only trivial dynamics in the linearly stable regime, which includes the general dynamics of pullers.

Experimental \cite{WuDiffusion1999,Yodh2007,Gollub2dbiomixing2011,Arratia2016} and particle-based computational realizations
\cite{UnderhillPoF2011,Koch2014diffusion,SubramanianRotationDiff2015,Najafi2015,morozov2014enhanced,Shelley2007} in the stable regime clearly demonstrate nontrivial flow and enhanced mixing even though the activity is less organized and vigorous than in unstable pusher systems. Such nontrivial dynamics can only be described by including the finite particulate nature of the microswimmers. One theoretical approach is to analyze a sample pair of microswimmers to understand the role of particle interactions~\cite{Underhill2008,UnderhillPoF2011,lauga_dance_2012,gyrya_model_2010,Michelin2CoupledSwimmer2009} or a microswimmer and tracer to understand mixing in the dilute limit~\cite{Koch2014diffusion,ThiffeaultNonGuassian2015,Gollub2dbiomixing2011,ThiffeaultSquirmers2010}.  We will adopt the complementary approach of adding stochastic fluxes to the continuum field equations to represent the mesoscopic fluctuations.  In addition to being more amenable to theoretical analysis, continuum field models are typically much less computationally expensive than direct particle simulations~\citep{CatesABPDimensionality2014}, though at the price of coarse-graining some interaction details.

The approach of adding physically appropriate mesoscopic noise to the diffusion equation has been used previously for a thermal equilibrium bath by Dean \cite{Dean1996}, Yoshimori \cite{Das2013}, and Biroli \cite{JstatmechStoField2007}. Cates \cite{CatePRL2008,Solon2015} applied this concept to some systems of active Brownian particles without explicit hydrodynamic interactions. A related approach was used by Lau and Lubensky \cite{Lubensky2009,Yodh2007}  who incorporated stochastic fluxes in a hydrodynamic model for microswimmers as a basis for computing statistical fluctuations and enhanced tracer diffusion. In this article, we pursue similar objectives, but proceed instead from a mesoscopic stochastic equation which can be more transparently derived from an underlying particle-level model, and which can be viewed as a
stochastic variation of the mean-field kinetic equation of \cite{Saintillan2008}. Our results quantifying enhanced tracer diffusion are similar to, but not in complete agreement with, those obtained by \cite{Lubensky2009}.

\section{Model Equations}
Our model consists of a suspension of $N_p$ rod-like dipole microswimmer particles in an incompressible, Newtonian fluid occupying a two-dimensional periodic square of side length $L$. We assume a concentration large enough for a continuum approximation to be reasonable but not so large that near-field and steric effects should be taken into account. The phase space number density $\psi\left(\bm{x},\bm{n},t\right)$ of microswimmers with position $\bm{x}$ and the unit vector of orientation $\bm{n}$ at time $t$ is governed by a mesoscopic field kinetic equation. This equation can be derived from hydrodynamically coupled Langevin equations for the individual microswimmers, following the procedures described for other models~\cite{Dean1996, Solon2015, Chavanis2008}, with the result:
\begin{align}
\frac{\partial\psi}{\partial t} & =  -\bm{\nabla_{x}}\cdot\left(\bm{\dot{x}}\psi\right)-\bm{\nabla_{n}}\cdot\left(\bm{\dot{n}}\psi\right)+D_{t}\nabla^2_{x}\psi \nonumber \\
 &   +D_{r}\nabla^2_{n}\psi+\Xi (\bm{x},\bm{n},t) \label{eq:fpequation}\\
 \bm{\dot{x}}&=v\bm{n}+\bm{u} (\bm{x},t)\label{eq:spatialvelocity}\\
 \bm{\dot{n}}&=\left(\bm{I}-\bm{nn}\right)\cdot\left(\gamma\bm{E} (\bm{x},t)-\bm{\Omega} (\bm{x},t)\right)\cdot\bm{n}.\label{eq:angularvelocity}
\end{align}
Here $\bm{\dot{x}}$ is the velocity of the microswimmer, $\bm{\dot{n}}$ is the angular velocity, the noise term $\Xi$ describes the divergence of stochastic fluxes from translational and rotational diffusion \cite{Dean1996}, $v$ is the swimming speed of an isolated swimmer, $\bm{u}$ is the local fluid velocity, $D_{t}$ is the translational diffusivity, $D_{r}$ is the rotational diffusivity, $\bm{I}$ is the identity tensor, $\bm{E}$ is the rate of strain tensor, $\bm{\Omega}$ is the vorticity tensor, and $\gamma$ is a shape parameter. For the case of long slender organisms including flagella, the shape parameter is $\gamma \approx1$. The swimming and flow lead to an advective flux while the diffusion leads to a stochastic flux. The Laplacian terms describe the mean divergence of these stochastic fluxes, while $ \Xi $ has zero mean and represents the fluctuations about the mean.

In the stochastic kinetic theory model, the instantaneous noise term in two-dimensional physical space is given by \cite{Dean1996}:
\begin{equation}
\Xi dt=\bm{\nabla_{x}}\cdot\left(\sqrt{2D_{t}\psi}\ d\bm{U} \left(\bm{x},\theta,t\right) \right)+\partial_{\theta}\left(\sqrt{2D_{r}\psi}\ dV \left(\bm{x},\theta,t\right) \right).
\end{equation}
The variables $d\bm{U}$ and $dV$ are independent Brownian motion processes following a Gaussian distribution with zero mean. The spatial noise contains $x$ and $y$ cartesian components that are independent. The correlations are given by
\begin{equation}
\langle d{U_{x}}\left(\bm{x},\theta,t\right)d{U_{x}}\left(\bm{x}',\theta',t'\right)\rangle= \langle d{U_{y}}\left(\bm{x},\theta,t\right)d{U_{y}}\left(\bm{x}',\theta',t'\right)\rangle= \langle dV \left(\bm{x},\theta,t\right)dV \left(\bm{x}',\theta',t'\right)\rangle= \delta\left(\bm{x}-\bm{x}'\right) \delta\left(\theta-\theta'\right) \delta\left(t-t'\right) dt dt'
\end{equation}
and the stochastic differential equation is to be interpreted in the It\^o sense. This noise on microswimmers is generally considered to be dominated by nonthermal effects such as noise from the operation of the molecular motors driving the swimming~\cite{Lubensky20072,Lubensky2009,GompperReview2015,GoldsteinARFM2015}.

The fluid flow is obtained from the time-independent Stokes equation for an incompressible fluid
\begin{equation}
-\eta\nabla_{x}^{2}\bm{u}+\bm{\nabla_{x}}q=\bm{\nabla_{x}\cdot\Sigma} , 
\end{equation}
where $ \eta $ is the dynamic viscosity and $q$ is the pressure. The active stress is
\begin{equation}
\bm{\Sigma}=d\int\psi\left(\bm{nn}-\bm{I}/2\right)d\bm{n} ,
\end{equation}
where $ d $ is the stresslet coefficient of the force dipole, which is
positive for pullers and negative for pushers. By using the time-independent Stokes equation, we are assuming the time scale of stress fluctuations induced by the microswimmer dynamics is slow relative to viscous relaxation, which amounts roughly to $ D_t \ll \eta/\rho $, where $ \rho $ is the fluid density, and wavenumbers $k$ such that $k \gg v\rho/\eta$, $\sqrt{D_r\rho/\eta}$, $\sqrt{|d|N_p \rho}/(\eta L) $.

These field equations lead to nontrivial dynamics even when the phase space density fluctuates around the uniform isotropic state. If these fluctuations are small enough, the field equations can be linearized, which is the subject of the rest of this article. In two dimensions, we can express the orientation vector as $ \bm{n} = (\cos \theta,\sin \theta)$.  We nondimensionalize the system using the time scale $t_{c}=\eta L^{2}/(N_p\left|d\right|)$ \cite{Bozorgi2011} and the length scale $l_{c}=v t_c=v\eta L^{2}/(N_p\left|d\right|)$ and scale the dimensional phase space density by its mean value $ N_p/(2\pi L^2)$. The key dimensionless parameters are $ \Lnd = L/l_c $, $\Dtnd=D_{t}t_{c}/l_{c}^{2}$, $\Drnd=D_{r}t_{c}$.  For simplicity of notation, in the rest of the article all other variables are nondimensional unless otherwise stated.

The scaled phase space density in a spectral expansion decouples the spatial Fourier modes for each spatial wavevector $\bm{k} =(k_x,k_y)$. In the spectral (Fourier) expansion, the phase space density is
\begin{equation}
\psi (\bm{x},\bm{n},t) =
\sum_{\bm{k} \in 2\pi \Lnd^{-1} \mathbb{Z}^2} \sum_{l=-\infty}^{\infty} \hpsi_{\bm{k},l} \expe^{\mathi (\bm{k}\cdot\bm{x}+l\theta)} .
\end{equation}
The Fourier coefficients for different angular modes $l$ are coupled and can be written in matrix form as a vector $\hbpsi_{\bm{k}} $ with components $\hpsi_{\bm{k},l}$ for $l=\ldots,-1,0,1,\ldots $. The evolution equation is
\begin{equation}\label{eq:PsiEvolution}
\frac{\partial \hbpsi_{\bm{k}}}{\partial t} =\mathbf{L_{\bm{k}}} \hbpsi +\mathi k_x\sqrt{\frac{2\Dtnd}{N_p}}\frac{d\mathbf{{U}}_{x,\bm{k}}}{dt} + \mathi k_y\sqrt{\frac{2\Dtnd}{N_p}}\frac{d\mathbf{{U}}_{y,\bm{k}}}{dt} +\mathi\sqrt{\frac{2\Drnd}{N_p}}\mathbf{{R}}_{\bm{k}}\frac{d\mathbf {V}_{\bm{k}}}{dt} ,
\end{equation}
where $\mathbf{R}_{\bm{k}}$ is diagonal with components $ R_{\bm{k},l,l'}=l \delta_{l,l'}$, while
the matrix $\mathbf{L}_{\bm{k}} $ is an infinite matrix with
components given, for $ \bm{k} \neq \bm{0} $,  by
\begin{align}
\begin{split}\label{eq:Ldefn}
\text{L}_{\bm{k},l,l'}=&-\frac{\mathi}{2} k \expe^{-\mathi \theta_k} \delta_{l,l'-1}-\frac{\mathi}{2} k \expe^{\mathi \theta_k} \delta_{l,l'+1}-\Dtnd k^{2}\delta_{l,l'} -\Drnd l^{2}\delta_{l,l'}
\\&+\delta_{l,2}\frac{p\gamma}{8}\left(-\delta_{l',2}+\expe^{-\mathi4\theta_k} \delta_{l',-2}\right)+\delta_{l,-2}\frac{p\gamma}{8}\left(\delta_{l',2}\expe^{\mathi4\theta_k} -\delta_{l',-2}\right)
\end{split}
\end{align}
where $\theta_k$ is the angle between the $x$ axis and $ \bm{k} $, and $p=d/|d|$ is the sign of the stresslet strength.

The noise terms are also written in Fourier space and collected in vectors $d\mathbf{{U}}_{x,\bm{k}}$ , $d\mathbf{{U}}_{y,\bm{k}}$ and $d\mathbf {V}_{\bm{k}}$ that are proportional to the Fourier transform of the physical space noise. These vectors have components $d\text{U}_{x,\bm{k},l}$, $d\text{U}_{y,\bm{k},l}$ and $d\text{V}_{\bm{k},l}$. All components are independent differentials of standard complex-valued Brownian motion, namely, Gaussian random functions with zero means and correlation functions given by
\begin{equation}
\langle d\text{U}^*_{x,\bm{k},l}(t) d\text {U}_{x,\bm{k},l}(t')\rangle = \langle d\text{U}^*_{y,\bm{k},l} (t) d\text {U}_{y,\bm{k},l} (t')\rangle = \langle d\text{V}^*_{\bm{k},l} (t) d\text{V}_{\bm{k},l} (t') \rangle =\delta\left(t-t'\right)dtdt'
\end{equation}

The linear evolution equation for $\hbpsi_{\bm{k}}$ for each $\bm{k}$ contains all the information about how the active motion and hydrodynamic interactions of the microswimmers lead to coordinated motion provided the phase space density remains close enough to the uniform isotropic state that the linear approximation is valid.   Within this approximation, we will focus on the properties of the long-time statistically stationary state (or non-equilibrium steady state (NESS)) of the
fluctuations and whose Fourier space amplitude will be proportional to $ N_p^{-1/2} $.

\section{Microswimmer fluctuations and correlations}
\begin{figure}
\includegraphics[width=0.45\textwidth]{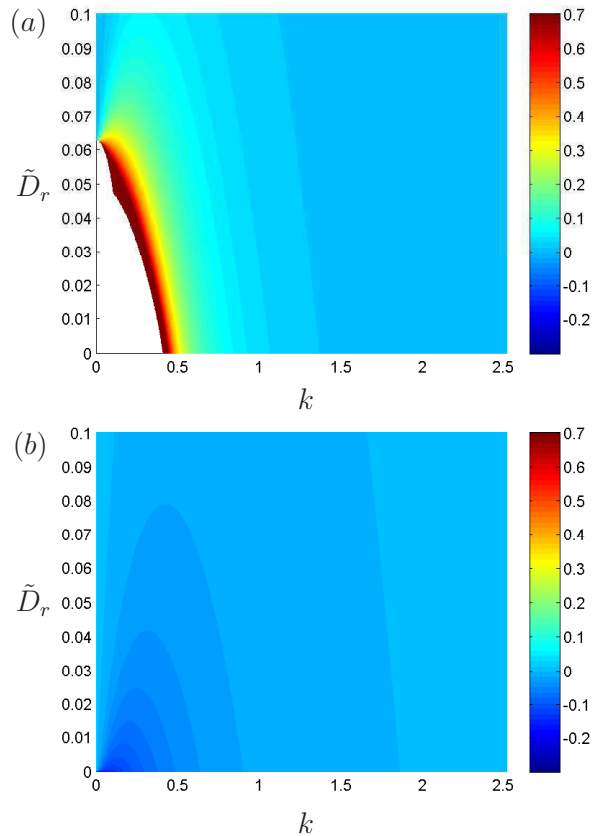}%
\caption{Fourier mode amplitudes of orientation correlation function, $\mbox{log}_{10} (N_p \hat{C}_{o,\bm{k}})$, for pushers (a) and pullers (b) as a function of wavenumber $k$ and nondimensional rotational diffusivity $\Drnd$. The blank region in panel (a)  corresponds to the region of linear instability of the pusher system. For both cases, the microswimmers are rod-like and slender ($\gamma=1$) with translational diffusivity of $\Dtnd=0.45$.}\label{Orientationpusherpuller}
\end{figure}

One important measure of group behavior is the local orientation of microswimmers, which for our nondimensionalization is 
\begin{equation}
\bm{N}\left(\bm{x},t\right)= \frac{1}{2 \pi} \int\bm{n}\psi\left(\bm{x},\bm{n}\right)d\bm{n} .
\end{equation}
Its spatial correlation function is defined as 
\begin{equation}
C_{o}\left(\bm{x}'\right)=\left\langle\bm{N}\left(\bm{x},t\right)\cdot\bm{N}\left(\bm{x}+\bm{x}',t\right)\right\rangle ,
\end{equation}
with its Fourier spectrum given by 
\begin{equation}
\hat{C}_{o,\bm{k}}=\frac{1}{\Lnd^2} \int_{0\times0}^{\Lnd\times \Lnd} C_{o}\left(\bm{x}'\right)\exp\left(-\mathi \bm{k}\cdot \bm{x}'\right)d\bm{x}' .
\end{equation}
Figure~\ref{Orientationpusherpuller} shows the spectrum from truncation of Eq.~\ref{eq:PsiEvolution} to components $ |l|\leq 32$, and solving the deterministic Lyapunov matrix equation characterizing the NESS statistics.

At large $\Drnd$, the spectrum of the orientation fluctuations approaches  the flat profile $ \hat{C}_{o,\bm{k}} = N_p^{-1} $ corresponding to independent microswimmers while for sufficiently small $ \Drnd $ and wavenumber $k $, the linearized theory is inapplicable for pushers due to the instability of the uniform isotropic state  \cite{Ramaswamy2002,Saintillan2008}.  At intermediate $ \Drnd $, the fluctuations in the orientation correlation field are enhanced for pushers and depressed for pullers, corresponding to their respective aligning and anti-aligning tendencies \cite{Shelley2007,PowersReview2009}. This can be understood by noting the polar orientation correlations can only be induced from the symmetric force dipole flows through coupling to swimming (first two terms $ \propto k $ in Eq.~\ref{eq:Ldefn}).  Translational diffusion destroys correlations at larger $k $.

To obtain further insight, we developed a ``slow swimming approximation'' in which the first two terms representing swimming in Eq.~\ref{eq:Ldefn} are perturbations to the remaining terms involving diffusion and active stress.  This approximation is formally valid for wavenumbers $ k \ll \min(\Drnd,4\Drnd+p\gamma/4) $ or $ k \gg \Dtnd^{-1} $, provided a singular perturbation expansion is developed to handle the degeneracy of rotational diffusion in the $ l=0 $ angular mode. Under the slow swimming approximation, the orientation fluctuation spectrum is
\begin{equation}
N_p \hat{C}_{o,\bm{k}}=1-\frac{p \gamma k^2/640}{\left(\Drnd+ \frac{k^{2}\Dtnd}{4}+\frac{p\gamma}{16}\right) \left(\Drnd+\frac{2k^{2}\Dtnd}{5}+\frac{p\gamma}{20}\right) \left(\Drnd+k^{2}\Dtnd\right) }+\cdots .
\end{equation}

\begin{figure}
\includegraphics[width=0.45\textwidth]{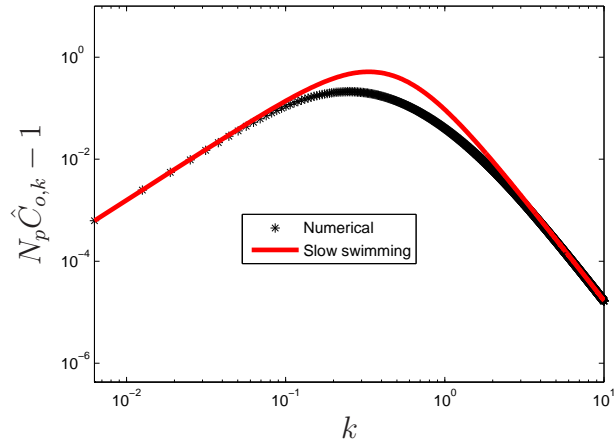}
\caption{Comparison of $N_p \hat{C}_{o,\bm{k}}-1$ between the numerical results from Fig.~\ref{Orientationpusherpuller}(a) and the slow swimming perturbation approximation. The translational diffusivity is $\Dtnd=0.45$, and the rotational diffusivity is $\Drnd=0.09$.}\label{fixDrocpur}
\end{figure}

Figure~\ref{fixDrocpur} compares the slow swimming approximation for slender pushers ($p\gamma=-1$) with $\Drnd=0.09$ and $\Dtnd=0.45$ against the results from Fig.~\ref{Orientationpusherpuller}.  We see agreement with both the $ k^2 $ scaling due to swimming at small $ k $ and $ k^{-4} $ scaling at large $k $ from translational diffusion.  For $ \Drnd \gg \gamma/16 $, the correlations peak at a wavenumber $ k_p \sim (\Dtnd/\Drnd)^{-1/2} $, where translational diffusion and rotational diffusion balance.  For pushers ($p=-1$), the hydrodynamic interactions cause the correlation length to increase as $\Drnd$ decreases and to diverge to the system size at the stability limit $ \Drnd = \gamma/16 $ of the uniform isotropic state.

In addition to orientation correlations, another important measure of fluctuations is the concentration of organisms relative to the mean concentration. Fluctuations in the spatial concentration defined as
\begin{equation}
C\left(\bm{x},t\right)= \frac{1}{2 \pi} \int\psi\left(\bm{x},\bm{n},t\right)d\bm{n}
\end{equation}
in our nondimensional scaling are found to be negligibly small in the linearized model. This is supported by the slow-swimming perturbation expansion which shows that the spatial autocorrelation function of the concentration fluctuations is delta-correlated in real space to the first three orders.  Recall that it is typically nonlinear effects or near-field effects that are responsible for the substantial concentration fluctuations seen in other studies \cite{Cate2DNonlinear2011,KlappNonlinear2011,Bozorgi20142,Marchetti_2010,CatesCluster2014}, while near-field effects are neglected here.

\begin{figure}
\includegraphics[width=3in]{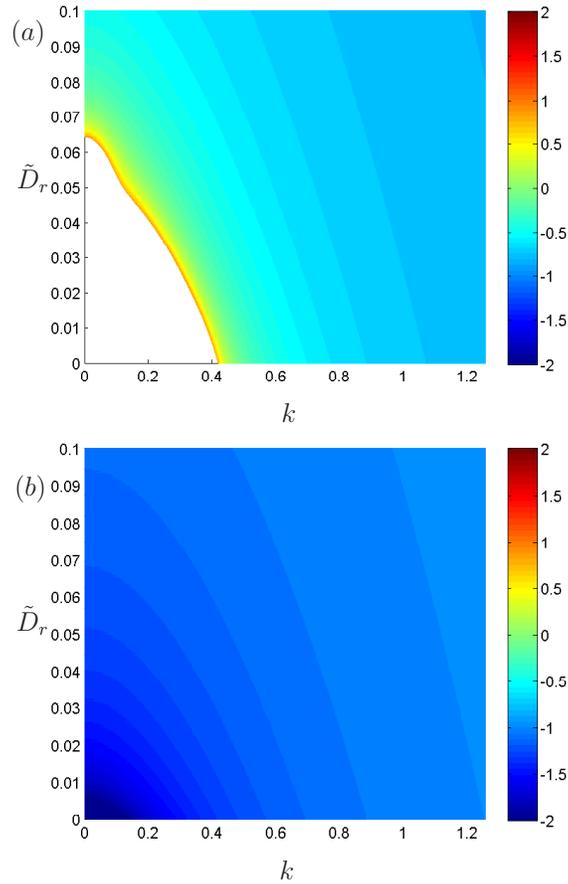}%
\caption{Fourier mode amplitudes of active stress correlation function, $\mbox{log}_{10} (N_p\hat{C}_{s,\bm{k}}|_{\theta_k=0})$, for pushers (a) and pullers (b) as a function of wavenumber $k$ and nondimensional rotational diffusivity $\Drnd$. The blank region in panel (a)  corresponds to the region of linear instability of the pusher system. For both cases, the microswimmers are rod-like and slender ($\gamma=1$) with translational diffusivity of $\Dtnd=0.45$.}\label{Stresspusherpuller}
\end{figure}

We consider next the shear component of the active stress, which under the deterministic mean field theory is the driver of the instability of the uniform isotropic state.  The results will be reported for the dimensional stress scaled by  $\eta/t_c=N_p|d|/L^2$, so the dimensionless stress is given by 
\begin{equation}
\bm{\Sigma}=\frac{p}{2 \pi}\int\psi\left(\bm{nn}-\bm{I}/2\right) d\bm{n} . 
\end{equation}
The shear stress correlation function in the non-equilibrium steady state (NESS) is 
\begin{equation}
C_{s}\left(\bm{x}'\right)=\left\langle \bm{\Sigma}_{xy}\left(\bm{x},t\right)\bm{\Sigma}_{xy}\left(\bm{x}+\bm{x}',t\right)\right\rangle . \end{equation}
This choice of shear component leads to a stress correlation that depends on $\theta_k$, so we will focus here on $\theta_k=0$. Similar to our approach with orientation correlations, the Fourier spectrum was calculated numerically and is shown in Figure \ref{Stresspusherpuller}. As with the orientation spectrum, the independent swimmer fluctuations are obtained for large enough $\Drnd$ or $k$.
The spectrum is monotonically varying in $k $, with the largest value for pushers at small $k$ and the smallest value for pullers as $k \to 0$. This is consistent with the mechanism of linear instability found in the mean field theory. The flows created by pushers enhance perturbations while the flows created by pullers suppress perturbations. The magnitude of this enhancement is largest for small $k$.

\begin{figure}
\includegraphics[width=3in]{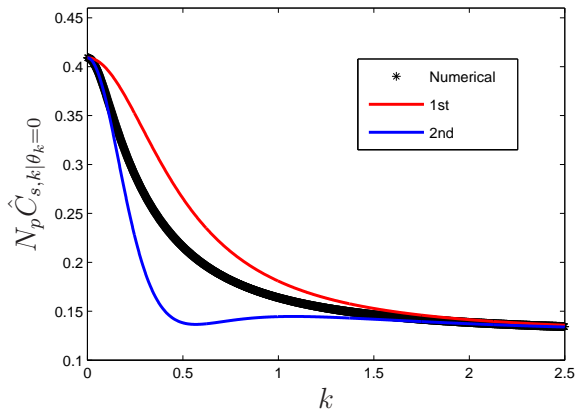}
\caption{Comparison of $N_{p} \hat{C}_{s,\bm{k}}|_{\theta_k=0}$ between the numerical results from Figure~\ref{Stresspusherpuller}(a) and the slow swimming perturbation approximation. The translational diffusivity is $\Dtnd=0.45$, and the rotational diffusivity is $\Drnd=0.09$. The red line uses the first approximation (first term) from equation \ref{eq:StressPert}, while the blue line uses both terms shown in equation \ref{eq:StressPert}. }\label{CompareStressPert}
\end{figure}

Many of the key features are illustrated by the perturbation expansion of the shear stress fluctuation spectrum under the slow swimming approximation, which is given by
\begin{align}
	\begin{split}\label{eq:StressPert}
	N_{p} & C_{s,\bm{k}}|_{\theta_k=0}	= \left(\frac{4\Drnd+k^{2}\Dtnd}{32\Drnd+8k^{2}\Dtnd+2p\gamma}\right)\\
	&\left[1+\frac{2p\gamma k^{2}\left(36\Drnd+8k^{2}\Dtnd+p\gamma\right)}{\left(52\Drnd+8k^{2}\Dtnd+p\gamma\right)\left(20\Drnd+8k^{2}\Dtnd+p\gamma\right)\left(16\Drnd+4k^{2}\Dtnd+p\gamma\right)\left(4\Drnd+k^{2}\Dtnd\right)}+\dots\right] .
	\end{split}
\end{align}
Figure \ref{CompareStressPert} compares the slow swimming approximation for slender pushers ($p\gamma=-1$) with $\Drnd=0.09$ and $\Dtnd=0.45$ against the results from Figure~\ref{Stresspusherpuller}. The leading order term of the approximation matches qualitatively with the numerical results. The slow swimming approximation is formally valid for wavenumbers $ k \ll \min(\Drnd,4\Drnd+p\gamma/4) $ or $ k \gg \Dtnd^{-1} $. For these values of $k$, using more terms in the expansion improves the comparison to the numerical results.

\section{Tracer dynamics}

We next consider how the statistical dynamics of the microswimmers enhance diffusion of passive objects (tracers) suspended in the fluid.
To quantify this diffusion, we have numerically integrated the stochastic equations of motion for the fluid velocity fluctuations, using the linearized equations together with non-Brownian point tracers advected by the local fluid velocity. The linearized version of Eqs.~\ref{eq:fpequation}-\ref{eq:angularvelocity} are discretized on a mesh with central finite difference derivatives in position or angle and explicit first order time stepping. The mesh contains $64$ elements each in $x$ and $y$ and $128$ in $\theta$. The timestep is $\Delta t=10^{-2}$. At each step, the active stress is evaluated on the mesh and the fluid velocity is found on the mesh from the Stokes equation using the FFTW algorithm. All of the results shown here have parameters $\Dtnd=0.45$, $\Lnd=50$, and $N_p=10^9$. The initial $200$ units of dimensionless time of the simulations are ignored to allow the system to reach the NESS.

\begin{figure}
\includegraphics[width=0.45\textwidth]{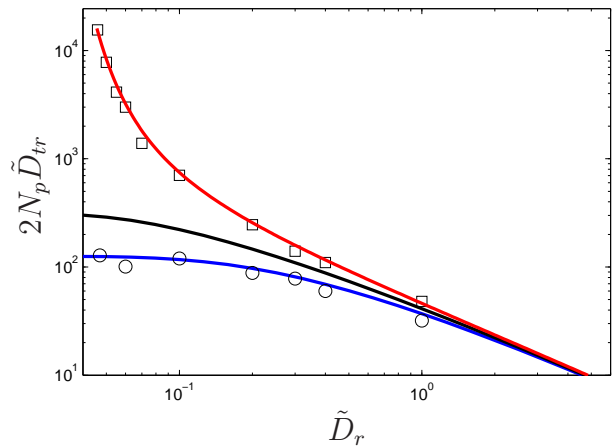}
\caption{The diffusivity $\tilde{D}_{tr}$ of non-Brownian passive point-particle tracers as a function of the rotational diffusivity of the microswimmer $\Drnd$. The symbols are from PDE-based stochastic simulations for slender rod-like pushers (squares) and pullers (circles). The lines are the solution to deterministic matrix equations that assume zero Kubo number. The solid red line is the calculation result for pushers, while the solid blue line is for pullers, and the solid black line is for non-interacting microswimmers.}
\label{matrixvspde}
\end{figure}

The mean-squared-displacement of the tracers is calculated and a diffusivity $\tilde{D}_{tr}$ is extracted from the slope at long time. These diffusivities are shown as symbols in Fig.~\ref{matrixvspde} for suspensions of slender rod-like ($\gamma=1$) pushers and pullers. At large $\Drnd$, the interactions between organisms do not significantly alter the swimmer dynamics, which leads to tracer diffusion that is the same for pushers and pullers. At smaller $\Drnd$, the interactions between organisms can alter the effective diffusion of the tracers. The tracer diffusivity is the integral of its Lagrangian velocity correlation function, which is generally difficult to calculate analytically.  However, the cases simulated in Fig.~\ref{matrixvspde} have a small Kubo number $Ku=v_f \tau_f/l_f \sim 10^{-5}$, where $v_f$ is the root-mean-square fluid velocity, $\tau_f$ is the correlation time of the fluid velocity, and $l_f$ is the correlation length of the fluid velocity~\cite{rk:sle}.  For small Kubo number, the Lagrangian velocity correlation function can be well approximated by the Eulerian one.

From the linearized stochastic equations (Eqs.~\ref{eq:PsiEvolution} and \ref{eq:Ldefn}), we derived deterministic (truncated) matrix equations for the Eulerian velocity correlation function, and solve these numerically to obtain small Kubo number approximations for the tracer diffusivity, which
are shown as solid curves in Fig.~\ref{matrixvspde}. Because the Kubo
number of the stochastic simulations are very small, they match well
with the small Kubo number approximation.  If all wavenumbers can be
treated under the slow swimming approximation $ (\Dtnd^{-1} \ll \min(\Drnd,4 \Drnd
+p\gamma/4))$, and we have the further condition $ {\Lnd^2}{\Dtnd^{-1}} (4 \Drnd
+p\gamma/4)\gg 1$ so that the sum over modes can be well approximated
  by an integral,  we can obtain an analytical expression for the tracer
diffusivity as
\begin{equation}
  \tilde{D}_{\mathrm{tr}} =  \frac{\Lnd^2}{64\pi N_p (4\Drnd +\frac{p\gamma}{4})^2} \left[\frac{p\gamma}{4} + 4\Drnd \left(\delta +\ln
    \left(\frac{\Lnd^2}{4\pi\Dtnd}(4 \Drnd +\frac{p\gamma}{4})\right)\right)\right] ,
  \end{equation}
where $ \delta \approx 1.8 $ is the Masser-Gramain constant. In dimensional terms this expression is
\begin{equation}
D_{\mathrm{tr}} =  \frac{N_p |d|^2}{256\pi \eta^2 D_r L^2 (1 +\frac{N_p d \gamma}{16 D_r \eta L^2})^2} \left[\frac{N_p d \gamma}{16 D_r \eta L^2} + \delta +\ln \left(\frac{L^2 D_r}{\pi D_t }(1 +\frac{N_p d\gamma}{16D_r \eta L^2})\right)\right] .
\end{equation}
For small concentrations the tracer diffusivity is proportional to the concentration of microswimmers and is the same for pushers and pullers. For higher concentrations, the hydrodynamic interactions of the microswimmers alter the tracer diffusion. This formula provides an analytical approximation of this effect.

Many experimental and simulation studies of
passive tracers in active systems show super-diffusion at shorter
times
\cite{GoldsteinDiffusion2009,morozov2014enhanced,caspi2000enhanced,Gollub2dbiomixing2011}. Yodh and coworkers~ \cite{Yodh2007}  found the
mean-squared-displacement scaled as $T^{3/2}$, where $T$ is the lag
time, and a fluctuating field model was used to deduce
this scaling \cite{Yodh2007,Lubensky2009}. The theory developed here also
produces super-diffusive scaling (nearly ballistic) of tracer motion for $T \ll \left(4D_r +p\gamma/4\right)^{-1}$ in two spatial dimensions. In particular, the dimensionless mean-squared displacement is approximately
\begin{equation}
\frac{\Lnd^2 T^2}{32 \pi N_p} \left[c + \ln \left(\frac{1}{(4\Drnd
    +\frac{p\gamma}{4})T}\right)\right]
\end{equation}
with
\begin{equation}
  c = 4-\tilde{\gamma} + \frac{4\Drnd}{4\Drnd +\frac{p\gamma}{4}}\left[\delta +
  \ln \left(\frac{\Lnd^2 (4\Drnd + \frac{p\gamma}{4})}{4\pi \Dtnd}\right)\right]
\end{equation}
where $ \tilde{\gamma} \approx 0.58 $ is the Euler-Mascheroni constant.

To better compare with experimental studies performed in three dimensions, we can examine the predictions of the fluctuating field theory in three dimensions. The predictions are most easily calculated analytically if we restrict attention to microswimmers which are not interacting. The long-time diffusion for non-Brownian
tracers moving in these flows can be computed analytically in three
dimensions under the assumptions of  low Kubo number, weak swimming, and $ \Dtnd \Lnd^{-2} \ll \Drnd$ (the rotational diffusion is faster than translational diffusion across the domain).  For these assumptions the long-time diffusivity is
\begin{equation}
\tilde{D}_{\mathrm{tr}} = \frac{\Lnd^3}{90 \pi \sqrt{6} N_p \Drnd^{1/2} \Dtnd^{1/2}} . 
\end{equation}
In dimensional form, this tracer diffusivity is
\begin{equation}
D_{\mathrm{tr}} = \frac{N_p |d|^2}{90 \pi \sqrt{6} L^3 \eta^2 D_{\mathrm{r}}^{1/2} D_{\mathrm{t}}^{1/2}}.
\end{equation}
The dimensional formula shows explicitly that the diffusivity is proportional to the microswimmer concentration and is independent of the sign of the dipole moment. This matches with previous theoretical work \cite{Underhill2008,UnderhillPoF2011,Koch2014diffusion}. For the low Kubo number calculation to be self-consistent, it is sufficient that $D_{\mathrm{tr}} \ll D_{\mathrm{t}}$. Using the expression above, this condition is equivalent to
\begin{equation}
\frac{N_p}{L^3} \ll \frac{\eta^2 D_{\mathrm{t}}^{3/2} D_{\mathrm{r}}^{1/2}}{|d|^2} .
\end{equation}
In dimensional terms, the point dipole approximation used throughout the article is valid when the swimmer size satisfies $l_{sw} \ll \sqrt{D_t/D_r}$. The slow swimming approximation corresponds to $v \ll \sqrt{D_r D_t}$. Our result matches exactly, including prefactors, with the slow swimming limiting behavior of the work of~\cite{Koch2014diffusion}.  Relative to~\cite{Koch2014diffusion}, our approach can incorporate hydrodynamic interactions of swimmers and describe the mean-square displacement of the tracers at all times in addition to the long-time diffusivity.

Indeed, for lag times $ T \ll 1/\Drnd $, with the same assumptions as above, we find a superdiffusive regime with the tracer mean-square displacement behaving as $4\Lnd^3 T^{3/2}/(45 N_p \pi^{3/2} \Dtnd^{1/2}) $,
which is consistent with previous experimental~\cite{caspi2000enhanced,Yodh2007} and theoretical~\cite{Lubensky2009} results. That is, super-diffusion of tracer motion in a three-dimensional flow
emerges already from the long-range spatial correlations in the fluid
generated by independently diffusing force dipoles, apart from
any hydrodynamic interactions influencing the dynamics of those force
dipoles, and does not seem to require a nonlinear theory as in~\cite{Lubensky2009} to obtain.

\section{Conclusion}

We have developed a stochastic kinetic theory that allows us
to better understand how hydrodynamic interactions alter the behavior
of suspensions of active particles. The equations can be used both when microswimmers act nearly
independently and when they produce large scale group behavior. The
results here have focused on situations where the uniform, isotropic
state is linearly stable, and the correlations and fluctuations for
the linearized dynamics have been quantified. Even within this regime, we have shown how
microswimmer interactions alter the correlations and
enhance the diffusion of passive tracers. This statistical structure cannot be
obtained  from a deterministic mean-field theory. Finally, we have shown that a simplified field theory produces the same super-diffusive scaling as observed in experiments. The stochastic kinetic theory can be used further to understand the role of fluctuations when the nonlinear terms are important and to understand the enhanced diffusion of tracers when the Kubo number is not small.

\section*{Acknowledgments}
We gratefully acknowledge support from NSF Grant No.~DMS-1211665.  PRK would like to thank the Isaac Newton Institute for Mathematical Sciences, Cambridge, for support and hospitality during the Stochastic Dynamical Systems in Biology programme, where some work on this paper was undertaken (EPSRC grant no EP/K032208/1.)


\begin{thebibliography}{60}%
\makeatletter
\providecommand \@ifxundefined [1]{%
 \@ifx{#1\undefined}
}%
\providecommand \@ifnum [1]{%
 \ifnum #1\expandafter \@firstoftwo
 \else \expandafter \@secondoftwo
 \fi
}%
\providecommand \@ifx [1]{%
 \ifx #1\expandafter \@firstoftwo
 \else \expandafter \@secondoftwo
 \fi
}%
\providecommand \natexlab [1]{#1}%
\providecommand \enquote  [1]{``#1''}%
\providecommand \bibnamefont  [1]{#1}%
\providecommand \bibfnamefont [1]{#1}%
\providecommand \citenamefont [1]{#1}%
\providecommand \href@noop [0]{\@secondoftwo}%
\providecommand \href [0]{\begingroup \@sanitize@url \@href}%
\providecommand \@href[1]{\@@startlink{#1}\@@href}%
\providecommand \@@href[1]{\endgroup#1\@@endlink}%
\providecommand \@sanitize@url [0]{\catcode `\\12\catcode `\$12\catcode
  `\&12\catcode `\#12\catcode `\^12\catcode `\_12\catcode `\%12\relax}%
\providecommand \@@startlink[1]{}%
\providecommand \@@endlink[0]{}%
\providecommand \url  [0]{\begingroup\@sanitize@url \@url }%
\providecommand \@url [1]{\endgroup\@href {#1}{\urlprefix }}%
\providecommand \urlprefix  [0]{URL }%
\providecommand \Eprint [0]{\href }%
\providecommand \doibase [0]{http://dx.doi.org/}%
\providecommand \selectlanguage [0]{\@gobble}%
\providecommand \bibinfo  [0]{\@secondoftwo}%
\providecommand \bibfield  [0]{\@secondoftwo}%
\providecommand \translation [1]{[#1]}%
\providecommand \BibitemOpen [0]{}%
\providecommand \bibitemStop [0]{}%
\providecommand \bibitemNoStop [0]{.\EOS\space}%
\providecommand \EOS [0]{\spacefactor3000\relax}%
\providecommand \BibitemShut  [1]{\csname bibitem#1\endcsname}%
\let\auto@bib@innerbib\@empty
\bibitem [{\citenamefont {Berg}(2004)}]{Bergbook}%
  \BibitemOpen
  \bibfield  {author} {\bibinfo {author} {\bibfnamefont {H.~C.}\ \bibnamefont
  {Berg}},\ }\href@noop {} {\emph {\bibinfo {title} {E. coli in Motion}}}\
  (\bibinfo  {publisher} {Springer, New York},\ \bibinfo {year}
  {2004})\BibitemShut {NoStop}%
\bibitem [{\citenamefont {Berg}\ and\ \citenamefont
  {Anderson}(1973)}]{BergBacteriaSwim1973}%
  \BibitemOpen
  \bibfield  {author} {\bibinfo {author} {\bibfnamefont {H.~C.}\ \bibnamefont
  {Berg}}\ and\ \bibinfo {author} {\bibfnamefont {R.~A.}\ \bibnamefont
  {Anderson}},\ }\bibfield  {title} {\enquote {\bibinfo {title} {Bacteria swim
  by rotating their flagellar filaments},}\ }\href@noop {} {\bibfield
  {journal} {\bibinfo  {journal} {Nature}\ }\textbf {\bibinfo {volume} {245}},\
  \bibinfo {pages} {380} (\bibinfo {year} {1973})}\BibitemShut {NoStop}%
\bibitem [{\citenamefont
  {et~al}(2011)}]{yilin_wu_et_al_self-organization_2011}%
  \BibitemOpen
  \bibfield  {author} {\bibinfo {author} {\bibfnamefont {Yilin~Wu}\
  \bibnamefont {et~al}},\ }\bibfield  {title} {\enquote {\bibinfo {title}
  {Self-organization in bacterial swarming: lessons from myxobacteria},}\
  }\href@noop {} {\bibfield  {journal} {\bibinfo  {journal} {Physical Biology}\
  }\textbf {\bibinfo {volume} {8}},\ \bibinfo {pages} {055003} (\bibinfo {year}
  {2011})}\BibitemShut {NoStop}%
\bibitem [{\citenamefont {Koch}\ and\ \citenamefont
  {Subramanian}(2011)}]{koch_collective_2011}%
  \BibitemOpen
  \bibfield  {author} {\bibinfo {author} {\bibfnamefont {Donald~L.}\
  \bibnamefont {Koch}}\ and\ \bibinfo {author} {\bibfnamefont {Ganesh}\
  \bibnamefont {Subramanian}},\ }\bibfield  {title} {\enquote {\bibinfo {title}
  {Collective {Hydrodynamics} of {Swimming} {Microorganisms}: {Living}
  {Fluids}},}\ }\href@noop {} {\bibfield  {journal} {\bibinfo  {journal} {Annu.
  Rev. Fluid Mech.}\ }\textbf {\bibinfo {volume} {43}},\ \bibinfo {pages}
  {637--659} (\bibinfo {year} {2011})}\BibitemShut {NoStop}%
\bibitem [{\citenamefont {Pedley}(2010{\natexlab{a}})}]{Pedley2010}%
  \BibitemOpen
  \bibfield  {author} {\bibinfo {author} {\bibfnamefont {T.~J.}\ \bibnamefont
  {Pedley}},\ }\bibfield  {title} {\enquote {\bibinfo {title} {Collective
  behaviour of swimming micro-organisms},}\ }\href@noop {} {\bibfield
  {journal} {\bibinfo  {journal} {Experimental Mechanics}\ }\textbf {\bibinfo
  {volume} {50}},\ \bibinfo {pages} {1293--1301} (\bibinfo {year}
  {2010}{\natexlab{a}})}\BibitemShut {NoStop}%
\bibitem [{\citenamefont {Ramaswamy}(2010)}]{RamaswamyReview2010}%
  \BibitemOpen
  \bibfield  {author} {\bibinfo {author} {\bibfnamefont {Sriram}\ \bibnamefont
  {Ramaswamy}},\ }\bibfield  {title} {\enquote {\bibinfo {title} {The mechanics
  and statistics of active matter},}\ }\href@noop {} {\bibfield  {journal}
  {\bibinfo  {journal} {Annual Review of Condensed Matter Physics}\ }\textbf
  {\bibinfo {volume} {1}},\ \bibinfo {pages} {323--345} (\bibinfo {year}
  {2010})}\BibitemShut {NoStop}%
\bibitem [{\citenamefont {Lauga}\ and\ \citenamefont
  {Powers}(2009)}]{PowersReview2009}%
  \BibitemOpen
  \bibfield  {author} {\bibinfo {author} {\bibfnamefont {Eric}\ \bibnamefont
  {Lauga}}\ and\ \bibinfo {author} {\bibfnamefont {Thomas~R}\ \bibnamefont
  {Powers}},\ }\bibfield  {title} {\enquote {\bibinfo {title} {The
  hydrodynamics of swimming microorganisms},}\ }\href@noop {} {\bibfield
  {journal} {\bibinfo  {journal} {Rep. Prog. Phys.}\ }\textbf {\bibinfo
  {volume} {72}},\ \bibinfo {pages} {096601} (\bibinfo {year}
  {2009})}\BibitemShut {NoStop}%
\bibitem [{\citenamefont {Guasto}\ \emph {et~al.}(2011)\citenamefont {Guasto},
  \citenamefont {Rusconi},\ and\ \citenamefont {Stocker}}]{StockerReview2011}%
  \BibitemOpen
  \bibfield  {author} {\bibinfo {author} {\bibfnamefont {Jeffrey~S.}\
  \bibnamefont {Guasto}}, \bibinfo {author} {\bibfnamefont {Roberto}\
  \bibnamefont {Rusconi}}, \ and\ \bibinfo {author} {\bibfnamefont {Roman}\
  \bibnamefont {Stocker}},\ }\bibfield  {title} {\enquote {\bibinfo {title}
  {Fluid mechanics of planktonic microorganisms},}\ }\href@noop {} {\bibfield
  {journal} {\bibinfo  {journal} {Annual Review of Fluid Mechanics}\ }\textbf
  {\bibinfo {volume} {44}},\ \bibinfo {pages} {373--400} (\bibinfo {year}
  {2011})}\BibitemShut {NoStop}%
\bibitem [{\citenamefont {Goldstein}(2015)}]{GoldsteinARFM2015}%
  \BibitemOpen
  \bibfield  {author} {\bibinfo {author} {\bibfnamefont {Raymond~E.}\
  \bibnamefont {Goldstein}},\ }\bibfield  {title} {\enquote {\bibinfo {title}
  {Green {Algae} as {Model} {Organisms} for {Biological} {Fluid} {Dynamics}},}\
  }\href@noop {} {\bibfield  {journal} {\bibinfo  {journal} {Annual Review of
  Fluid Mechanics}\ }\textbf {\bibinfo {volume} {47}},\ \bibinfo {pages}
  {343--375} (\bibinfo {year} {2015})}\BibitemShut {NoStop}%
\bibitem [{\citenamefont {Krishnamurthy}\ and\ \citenamefont
  {Subramanian}(2015)}]{SubramanianRotationDiff2015}%
  \BibitemOpen
  \bibfield  {author} {\bibinfo {author} {\bibfnamefont {Deepak}\ \bibnamefont
  {Krishnamurthy}}\ and\ \bibinfo {author} {\bibfnamefont {Ganesh}\
  \bibnamefont {Subramanian}},\ }\bibfield  {title} {\enquote {\bibinfo {title}
  {Collective motion in a suspension of micro-swimmers that run-and-tumble and
  rotary diffuse},}\ }\href@noop {} {\bibfield  {journal} {\bibinfo  {journal}
  {Journal of Fluid Mechanics}\ }\textbf {\bibinfo {volume} {781}},\ \bibinfo
  {pages} {422--466} (\bibinfo {year} {2015})}\BibitemShut {NoStop}%
\bibitem [{\citenamefont {Marchetti}\ \emph {et~al.}(2013)\citenamefont
  {Marchetti}, \citenamefont {Joanny}, \citenamefont {Ramaswamy}, \citenamefont
  {Liverpool}, \citenamefont {Prost}, \citenamefont {Rao},\ and\ \citenamefont
  {Simha}}]{marchetti_hydrodynamics_2013}%
  \BibitemOpen
  \bibfield  {author} {\bibinfo {author} {\bibfnamefont {M.~C.}\ \bibnamefont
  {Marchetti}}, \bibinfo {author} {\bibfnamefont {J.~F.}\ \bibnamefont
  {Joanny}}, \bibinfo {author} {\bibfnamefont {S.}~\bibnamefont {Ramaswamy}},
  \bibinfo {author} {\bibfnamefont {T.~B.}\ \bibnamefont {Liverpool}}, \bibinfo
  {author} {\bibfnamefont {J.}~\bibnamefont {Prost}}, \bibinfo {author}
  {\bibfnamefont {Madan}\ \bibnamefont {Rao}}, \ and\ \bibinfo {author}
  {\bibfnamefont {R.~Aditi}\ \bibnamefont {Simha}},\ }\bibfield  {title}
  {\enquote {\bibinfo {title} {Hydrodynamics of soft active matter},}\
  }\href@noop {} {\bibfield  {journal} {\bibinfo  {journal} {Reviews of Modern
  Physics}\ }\textbf {\bibinfo {volume} {85}},\ \bibinfo {pages} {1143--1189}
  (\bibinfo {year} {2013})}\BibitemShut {NoStop}%
\bibitem [{\citenamefont {Cates}(2012)}]{CateQuestion2012}%
  \BibitemOpen
  \bibfield  {author} {\bibinfo {author} {\bibfnamefont {M~E}\ \bibnamefont
  {Cates}},\ }\bibfield  {title} {\enquote {\bibinfo {title} {Diffusive
  transport without detailed balance in motile bacteria: does microbiology need
  statistical physics?}}\ }\href@noop {} {\bibfield  {journal} {\bibinfo
  {journal} {Reports on Progress in Physics}\ }\textbf {\bibinfo {volume}
  {75}},\ \bibinfo {pages} {042601} (\bibinfo {year} {2012})}\BibitemShut
  {NoStop}%
\bibitem [{\citenamefont {Thompson}\ \emph {et~al.}(2011)\citenamefont
  {Thompson}, \citenamefont {Tailleur}, \citenamefont {Cates},\ and\
  \citenamefont {Blythe}}]{CateLatticeMeanfield2011}%
  \BibitemOpen
  \bibfield  {author} {\bibinfo {author} {\bibfnamefont {A~G}\ \bibnamefont
  {Thompson}}, \bibinfo {author} {\bibfnamefont {J}~\bibnamefont {Tailleur}},
  \bibinfo {author} {\bibfnamefont {M~E}\ \bibnamefont {Cates}}, \ and\
  \bibinfo {author} {\bibfnamefont {R~A}\ \bibnamefont {Blythe}},\ }\bibfield
  {title} {\enquote {\bibinfo {title} {Lattice models of nonequilibrium
  bacterial dynamics},}\ }\href@noop {} {\bibfield  {journal} {\bibinfo
  {journal} {Journal of Statistical Mechanics: Theory and Experiment}\ }\textbf
  {\bibinfo {volume} {2011}},\ \bibinfo {pages} {P02029} (\bibinfo {year}
  {2011})}\BibitemShut {NoStop}%
\bibitem [{\citenamefont {Elgeti}\ \emph {et~al.}(2015)\citenamefont {Elgeti},
  \citenamefont {Winkler},\ and\ \citenamefont {Gompper}}]{GompperReview2015}%
  \BibitemOpen
  \bibfield  {author} {\bibinfo {author} {\bibfnamefont {J}~\bibnamefont
  {Elgeti}}, \bibinfo {author} {\bibfnamefont {R~G}\ \bibnamefont {Winkler}}, \
  and\ \bibinfo {author} {\bibfnamefont {G}~\bibnamefont {Gompper}},\
  }\bibfield  {title} {\enquote {\bibinfo {title} {Physics of
  microswimmers---single particle motion and collective behavior: a review},}\
  }\href@noop {} {\bibfield  {journal} {\bibinfo  {journal} {Reports on
  Progress in Physics}\ }\textbf {\bibinfo {volume} {78}},\ \bibinfo {pages}
  {056601} (\bibinfo {year} {2015})}\BibitemShut {NoStop}%
\bibitem [{\citenamefont {Saintillan}\ and\ \citenamefont
  {Shelley}(2008)}]{Saintillan2008}%
  \BibitemOpen
  \bibfield  {author} {\bibinfo {author} {\bibfnamefont {David}\ \bibnamefont
  {Saintillan}}\ and\ \bibinfo {author} {\bibfnamefont {Michael~J.}\
  \bibnamefont {Shelley}},\ }\bibfield  {title} {\enquote {\bibinfo {title}
  {Instabilities and pattern formation in active particle suspensions: Kinetic
  theory and continuum simulations},}\ }\href@noop {} {\bibfield  {journal}
  {\bibinfo  {journal} {Phys. Rev. Lett.}\ }\textbf {\bibinfo {volume} {100}},\
  \bibinfo {pages} {178103} (\bibinfo {year} {2008})}\BibitemShut {NoStop}%
\bibitem [{\citenamefont {{Saintillan}}(2009)}]{SaintillanDilute2009}%
  \BibitemOpen
  \bibfield  {author} {\bibinfo {author} {\bibfnamefont {D.}~\bibnamefont
  {{Saintillan}}},\ }\bibfield  {title} {\enquote {\bibinfo {title} {The dilute
  rheology of swimming suspensions: A simple kinetic model},}\ }\href@noop {}
  {\bibfield  {journal} {\bibinfo  {journal} {Experimental Mechanics}\ }\textbf
  {\bibinfo {volume} {50}},\ \bibinfo {pages} {1275--1281} (\bibinfo {year}
  {2009})}\BibitemShut {NoStop}%
\bibitem [{\citenamefont {Saintillan}(2012)}]{Saintillanbook2012}%
  \BibitemOpen
  \bibfield  {author} {\bibinfo {author} {\bibfnamefont {David}\ \bibnamefont
  {Saintillan}},\ }\enquote {\bibinfo {title} {Natural locomotion in fluids and
  on surfaces: Swimming, flying, and sliding},}\ \ (\bibinfo  {publisher}
  {Springer New York},\ \bibinfo {address} {New York, NY},\ \bibinfo {year}
  {2012})\ Chap.\ \bibinfo {chapter} {Kinetic Models for Biologically Active
  Suspensions}, pp.\ \bibinfo {pages} {53--71}\BibitemShut {NoStop}%
\bibitem [{\citenamefont {Gross}\ \emph {et~al.}(2010)\citenamefont {Gross},
  \citenamefont {Adhikari}, \citenamefont {Cates},\ and\ \citenamefont
  {Varnik}}]{Cateslatticeboltzmann2010}%
  \BibitemOpen
  \bibfield  {author} {\bibinfo {author} {\bibfnamefont {M.}~\bibnamefont
  {Gross}}, \bibinfo {author} {\bibfnamefont {R.}~\bibnamefont {Adhikari}},
  \bibinfo {author} {\bibfnamefont {M.~E.}\ \bibnamefont {Cates}}, \ and\
  \bibinfo {author} {\bibfnamefont {F.}~\bibnamefont {Varnik}},\ }\bibfield
  {title} {\enquote {\bibinfo {title} {Thermal fluctuations in the lattice
  boltzmann method for nonideal fluids},}\ }\href@noop {} {\bibfield  {journal}
  {\bibinfo  {journal} {Phys. Rev. E}\ }\textbf {\bibinfo {volume} {82}},\
  \bibinfo {pages} {056714} (\bibinfo {year} {2010})}\BibitemShut {NoStop}%
\bibitem [{\citenamefont {Pedley}(2010{\natexlab{b}})}]{PEDLEYrevisit2010}%
  \BibitemOpen
  \bibfield  {author} {\bibinfo {author} {\bibfnamefont {T.~J.}\ \bibnamefont
  {Pedley}},\ }\bibfield  {title} {\enquote {\bibinfo {title} {Instability of
  uniform micro-organism suspensions revisited},}\ }\href@noop {} {\bibfield
  {journal} {\bibinfo  {journal} {Journal of Fluid Mechanics}\ }\textbf
  {\bibinfo {volume} {647}},\ \bibinfo {pages} {335--359} (\bibinfo {year}
  {2010}{\natexlab{b}})}\BibitemShut {NoStop}%
\bibitem [{\citenamefont {Pahlavan}\ and\ \citenamefont
  {Saintillan}(2011)}]{Staintillan2011}%
  \BibitemOpen
  \bibfield  {author} {\bibinfo {author} {\bibfnamefont {Amir~Alizadeh}\
  \bibnamefont {Pahlavan}}\ and\ \bibinfo {author} {\bibfnamefont {David}\
  \bibnamefont {Saintillan}},\ }\bibfield  {title} {\enquote {\bibinfo {title}
  {Instability regimes in flowing suspensions of swimming micro-organisms},}\
  }\href@noop {} {\bibfield  {journal} {\bibinfo  {journal} {Physics of
  Fluids}\ }\textbf {\bibinfo {volume} {23}},\ \bibinfo {pages} {011901}
  (\bibinfo {year} {2011})}\BibitemShut {NoStop}%
\bibitem [{\citenamefont {Saintillan}\ and\ \citenamefont
  {Shelley}(2012)}]{ShelleyJRSI2012}%
  \BibitemOpen
  \bibfield  {author} {\bibinfo {author} {\bibfnamefont {David}\ \bibnamefont
  {Saintillan}}\ and\ \bibinfo {author} {\bibfnamefont {Michael~J.}\
  \bibnamefont {Shelley}},\ }\bibfield  {title} {\enquote {\bibinfo {title}
  {Emergence of coherent structures and large-scale flows in motile
  suspensions},}\ }\href@noop {} {\bibfield  {journal} {\bibinfo  {journal} {J.
  R. Soc. Interface}\ }\textbf {\bibinfo {volume} {9}},\ \bibinfo {pages}
  {571--585} (\bibinfo {year} {2012})}\BibitemShut {NoStop}%
\bibitem [{\citenamefont {Mishra}\ \emph {et~al.}(2010)\citenamefont {Mishra},
  \citenamefont {Baskaran},\ and\ \citenamefont {Marchetti}}]{Marchetti2010}%
  \BibitemOpen
  \bibfield  {author} {\bibinfo {author} {\bibfnamefont {Shradha}\ \bibnamefont
  {Mishra}}, \bibinfo {author} {\bibfnamefont {Aparna}\ \bibnamefont
  {Baskaran}}, \ and\ \bibinfo {author} {\bibfnamefont {M.~Cristina}\
  \bibnamefont {Marchetti}},\ }\bibfield  {title} {\enquote {\bibinfo {title}
  {Fluctuations and pattern formation in self-propelled particles},}\
  }\href@noop {} {\bibfield  {journal} {\bibinfo  {journal} {Phys. Rev. E}\
  }\textbf {\bibinfo {volume} {81}},\ \bibinfo {pages} {061916} (\bibinfo
  {year} {2010})}\BibitemShut {NoStop}%
\bibitem [{\citenamefont {Simha}\ and\ \citenamefont
  {Ramaswamy}(2002)}]{Ramaswamy2002}%
  \BibitemOpen
  \bibfield  {author} {\bibinfo {author} {\bibfnamefont {R.~Aditi}\
  \bibnamefont {Simha}}\ and\ \bibinfo {author} {\bibfnamefont
  {S.}~\bibnamefont {Ramaswamy}},\ }\bibfield  {title} {\enquote {\bibinfo
  {title} {Hydrodynamic fluctuations and instabilities in ordered suspensions
  of self-propelled particles},}\ }\href@noop {} {\bibfield  {journal}
  {\bibinfo  {journal} {Physical Review Letters}\ }\textbf {\bibinfo {volume}
  {89}},\ \bibinfo {pages} {058101} (\bibinfo {year} {2002})}\BibitemShut
  {NoStop}%
\bibitem [{\citenamefont {Evans}\ \emph {et~al.}(2011)\citenamefont {Evans},
  \citenamefont {Ishikawa}, \citenamefont {Yamaguchi},\ and\ \citenamefont
  {Lauga}}]{LaugaDense2011}%
  \BibitemOpen
  \bibfield  {author} {\bibinfo {author} {\bibfnamefont {Arthur~A.}\
  \bibnamefont {Evans}}, \bibinfo {author} {\bibfnamefont {Takuji}\
  \bibnamefont {Ishikawa}}, \bibinfo {author} {\bibfnamefont {Takami}\
  \bibnamefont {Yamaguchi}}, \ and\ \bibinfo {author} {\bibfnamefont {Eric}\
  \bibnamefont {Lauga}},\ }\bibfield  {title} {\enquote {\bibinfo {title}
  {Orientational order in concentrated suspensions of spherical
  microswimmers},}\ }\href@noop {} {\bibfield  {journal} {\bibinfo  {journal}
  {Physics of Fluids}\ }\textbf {\bibinfo {volume} {23}} (\bibinfo {year}
  {2011})}\BibitemShut {NoStop}%
\bibitem [{\citenamefont {Fielding}\ \emph {et~al.}(2011)\citenamefont
  {Fielding}, \citenamefont {Marenduzzo},\ and\ \citenamefont
  {Cates}}]{Cate2DNonlinear2011}%
  \BibitemOpen
  \bibfield  {author} {\bibinfo {author} {\bibfnamefont {S.~M.}\ \bibnamefont
  {Fielding}}, \bibinfo {author} {\bibfnamefont {D.}~\bibnamefont
  {Marenduzzo}}, \ and\ \bibinfo {author} {\bibfnamefont {M.~E.}\ \bibnamefont
  {Cates}},\ }\bibfield  {title} {\enquote {\bibinfo {title} {Nonlinear
  dynamics and rheology of active fluids: Simulations in two dimensions},}\
  }\href@noop {} {\bibfield  {journal} {\bibinfo  {journal} {Phys. Rev. E}\
  }\textbf {\bibinfo {volume} {83}},\ \bibinfo {pages} {041910} (\bibinfo
  {year} {2011})}\BibitemShut {NoStop}%
\bibitem [{\citenamefont {Zhang}\ \emph {et~al.}(2010)\citenamefont {Zhang},
  \citenamefont {Be'er}, \citenamefont {Florin},\ and\ \citenamefont
  {Swinney}}]{SwinneyDenseExp2010}%
  \BibitemOpen
  \bibfield  {author} {\bibinfo {author} {\bibfnamefont {H.~P.}\ \bibnamefont
  {Zhang}}, \bibinfo {author} {\bibfnamefont {Avraham}\ \bibnamefont {Be'er}},
  \bibinfo {author} {\bibfnamefont {E.-L.}\ \bibnamefont {Florin}}, \ and\
  \bibinfo {author} {\bibfnamefont {Harry~L.}\ \bibnamefont {Swinney}},\
  }\bibfield  {title} {\enquote {\bibinfo {title} {Collective motion and
  density fluctuations in bacterial colonies},}\ }\href@noop {} {\bibfield
  {journal} {\bibinfo  {journal} {Proceedings of the National Academy of
  Sciences}\ }\textbf {\bibinfo {volume} {107}},\ \bibinfo {pages}
  {13626--13630} (\bibinfo {year} {2010})}\BibitemShut {NoStop}%
\bibitem [{\citenamefont {Wensink}\ and\ \citenamefont
  {L{\"o}wen}(2012)}]{LowenDense2012}%
  \BibitemOpen
  \bibfield  {author} {\bibinfo {author} {\bibfnamefont {H~H}\ \bibnamefont
  {Wensink}}\ and\ \bibinfo {author} {\bibfnamefont {H}~\bibnamefont
  {L{\"o}wen}},\ }\bibfield  {title} {\enquote {\bibinfo {title} {Emergent
  states in dense systems of active rods: from swarming to turbulence},}\
  }\href@noop {} {\bibfield  {journal} {\bibinfo  {journal} {Journal of
  Physics: Condensed Matter}\ }\textbf {\bibinfo {volume} {24}},\ \bibinfo
  {pages} {464130} (\bibinfo {year} {2012})}\BibitemShut {NoStop}%
\bibitem [{\citenamefont {Bozorgi}\ and\ \citenamefont
  {Underhill}(2014{\natexlab{a}})}]{Bozorgi2014}%
  \BibitemOpen
  \bibfield  {author} {\bibinfo {author} {\bibfnamefont {Yaser}\ \bibnamefont
  {Bozorgi}}\ and\ \bibinfo {author} {\bibfnamefont {Patrick~T.}\ \bibnamefont
  {Underhill}},\ }\bibfield  {title} {\enquote {\bibinfo {title}
  {Large-amplitude oscillatory shear rheology of dilute active suspensions},}\
  }\href@noop {} {\bibfield  {journal} {\bibinfo  {journal} {Rheologica Acta}\
  }\textbf {\bibinfo {volume} {53}},\ \bibinfo {pages} {899--909} (\bibinfo
  {year} {2014}{\natexlab{a}})}\BibitemShut {NoStop}%
\bibitem [{\citenamefont {Bozorgi}\ and\ \citenamefont
  {Underhill}(2014{\natexlab{b}})}]{Bozorgi20142}%
  \BibitemOpen
  \bibfield  {author} {\bibinfo {author} {\bibfnamefont {Yaser}\ \bibnamefont
  {Bozorgi}}\ and\ \bibinfo {author} {\bibfnamefont {Patrick~T.}\ \bibnamefont
  {Underhill}},\ }\bibfield  {title} {\enquote {\bibinfo {title} {Effects of
  elasticity on the nonlinear collective dynamics of self-propelled
  particles},}\ }\href@noop {} {\bibfield  {journal} {\bibinfo  {journal}
  {Journal of Non-Newtonian Fluid Mechanics}\ }\textbf {\bibinfo {volume}
  {214}},\ \bibinfo {pages} {69 -- 77} (\bibinfo {year}
  {2014}{\natexlab{b}})}\BibitemShut {NoStop}%
\bibitem [{\citenamefont {Wu}\ and\ \citenamefont
  {Libchaber}(2000)}]{WuDiffusion1999}%
  \BibitemOpen
  \bibfield  {author} {\bibinfo {author} {\bibfnamefont {Xiao-Lun}\
  \bibnamefont {Wu}}\ and\ \bibinfo {author} {\bibfnamefont {Albert}\
  \bibnamefont {Libchaber}},\ }\bibfield  {title} {\enquote {\bibinfo {title}
  {Particle diffusion in a quasi-two-dimensional bacterial bath},}\ }\href@noop
  {} {\bibfield  {journal} {\bibinfo  {journal} {Phys. Rev. Lett.}\ }\textbf
  {\bibinfo {volume} {84}},\ \bibinfo {pages} {3017--3020} (\bibinfo {year}
  {2000})}\BibitemShut {NoStop}%
\bibitem [{\citenamefont {Chen}\ \emph {et~al.}(2007)\citenamefont {Chen},
  \citenamefont {Lau}, \citenamefont {Hough}, \citenamefont {Islam},
  \citenamefont {Goulian}, \citenamefont {Lubensky},\ and\ \citenamefont
  {Yodh}}]{Yodh2007}%
  \BibitemOpen
  \bibfield  {author} {\bibinfo {author} {\bibfnamefont {D.~T.}\ \bibnamefont
  {Chen}}, \bibinfo {author} {\bibfnamefont {A.}~\bibnamefont {Lau}}, \bibinfo
  {author} {\bibfnamefont {L.~A.}\ \bibnamefont {Hough}}, \bibinfo {author}
  {\bibfnamefont {M.~F.}\ \bibnamefont {Islam}}, \bibinfo {author}
  {\bibfnamefont {M.}~\bibnamefont {Goulian}}, \bibinfo {author} {\bibfnamefont
  {T.~C.}\ \bibnamefont {Lubensky}}, \ and\ \bibinfo {author} {\bibfnamefont
  {A.~G.}\ \bibnamefont {Yodh}},\ }\bibfield  {title} {\enquote {\bibinfo
  {title} {Fluctuations and rheology in active bacterial suspensions},}\
  }\href@noop {} {\bibfield  {journal} {\bibinfo  {journal} {Physical Review
  letters}\ }\textbf {\bibinfo {volume} {99}},\ \bibinfo {pages} {148302}
  (\bibinfo {year} {2007})}\BibitemShut {NoStop}%
\bibitem [{\citenamefont {Kurtuldu}\ \emph {et~al.}(2011)\citenamefont
  {Kurtuldu}, \citenamefont {Guasto}, \citenamefont {Johnson},\ and\
  \citenamefont {Gollub}}]{Gollub2dbiomixing2011}%
  \BibitemOpen
  \bibfield  {author} {\bibinfo {author} {\bibfnamefont {H{\"u}seyin}\
  \bibnamefont {Kurtuldu}}, \bibinfo {author} {\bibfnamefont {Jeffrey~S.}\
  \bibnamefont {Guasto}}, \bibinfo {author} {\bibfnamefont {Karl~A.}\
  \bibnamefont {Johnson}}, \ and\ \bibinfo {author} {\bibfnamefont {J.~P.}\
  \bibnamefont {Gollub}},\ }\bibfield  {title} {\enquote {\bibinfo {title}
  {Enhancement of biomixing by swimming algal cells in two-dimensional
  films},}\ }\href@noop {} {\bibfield  {journal} {\bibinfo  {journal}
  {Proceedings of the National Academy of Sciences}\ }\textbf {\bibinfo
  {volume} {108}},\ \bibinfo {pages} {10391--10395} (\bibinfo {year}
  {2011})}\BibitemShut {NoStop}%
\bibitem [{\citenamefont {Patteson}\ \emph {et~al.}(2016)\citenamefont
  {Patteson}, \citenamefont {Gopinath}, \citenamefont {Purohit},\ and\
  \citenamefont {Arratia}}]{Arratia2016}%
  \BibitemOpen
  \bibfield  {author} {\bibinfo {author} {\bibfnamefont {Alison~E.}\
  \bibnamefont {Patteson}}, \bibinfo {author} {\bibfnamefont {Arvind}\
  \bibnamefont {Gopinath}}, \bibinfo {author} {\bibfnamefont {Prashant~K.}\
  \bibnamefont {Purohit}}, \ and\ \bibinfo {author} {\bibfnamefont {Paulo~E.}\
  \bibnamefont {Arratia}},\ }\bibfield  {title} {\enquote {\bibinfo {title}
  {Particle diffusion in active fluids is non-monotonic in size},}\ }\href@noop
  {} {\bibfield  {journal} {\bibinfo  {journal} {Soft Matter}\ }\textbf
  {\bibinfo {volume} {12}},\ \bibinfo {pages} {2365--2372} (\bibinfo {year}
  {2016})}\BibitemShut {NoStop}%
\bibitem [{\citenamefont {Underhill}\ and\ \citenamefont
  {Graham}(2011)}]{UnderhillPoF2011}%
  \BibitemOpen
  \bibfield  {author} {\bibinfo {author} {\bibfnamefont {Patrick~T.}\
  \bibnamefont {Underhill}}\ and\ \bibinfo {author} {\bibfnamefont
  {Michael~D.}\ \bibnamefont {Graham}},\ }\bibfield  {title} {\enquote
  {\bibinfo {title} {Correlations and fluctuations of stress and velocity in
  suspensions of swimming microorganisms},}\ }\href@noop {} {\bibfield
  {journal} {\bibinfo  {journal} {Physics of Fluids}\ }\textbf {\bibinfo
  {volume} {23}},\ \bibinfo {eid} {121902} (\bibinfo {year}
  {2011})}\BibitemShut {NoStop}%
\bibitem [{\citenamefont {Kasyap}\ \emph {et~al.}(2014)\citenamefont {Kasyap},
  \citenamefont {Koch},\ and\ \citenamefont {Wu}}]{Koch2014diffusion}%
  \BibitemOpen
  \bibfield  {author} {\bibinfo {author} {\bibfnamefont {TV}~\bibnamefont
  {Kasyap}}, \bibinfo {author} {\bibfnamefont {Donald~L}\ \bibnamefont {Koch}},
  \ and\ \bibinfo {author} {\bibfnamefont {Mingming}\ \bibnamefont {Wu}},\
  }\bibfield  {title} {\enquote {\bibinfo {title} {Hydrodynamic tracer
  diffusion in suspensions of swimming bacteria},}\ }\href@noop {} {\bibfield
  {journal} {\bibinfo  {journal} {Physics of Fluids (1994-present)}\ }\textbf
  {\bibinfo {volume} {26}},\ \bibinfo {pages} {081901} (\bibinfo {year}
  {2014})}\BibitemShut {NoStop}%
\bibitem [{\citenamefont {Moradi}\ and\ \citenamefont
  {Najafi}(2015)}]{Najafi2015}%
  \BibitemOpen
  \bibfield  {author} {\bibinfo {author} {\bibfnamefont {Moslem}\ \bibnamefont
  {Moradi}}\ and\ \bibinfo {author} {\bibfnamefont {Ali}\ \bibnamefont
  {Najafi}},\ }\bibfield  {title} {\enquote {\bibinfo {title} {Rheological
  properties of a dilute suspension of self-propelled particles},}\ }\href@noop
  {} {\bibfield  {journal} {\bibinfo  {journal} {EPL (Europhysics Letters)}\
  }\textbf {\bibinfo {volume} {109}},\ \bibinfo {pages} {24001} (\bibinfo
  {year} {2015})}\BibitemShut {NoStop}%
\bibitem [{\citenamefont {Morozov}\ and\ \citenamefont
  {Marenduzzo}(2014)}]{morozov2014enhanced}%
  \BibitemOpen
  \bibfield  {author} {\bibinfo {author} {\bibfnamefont {Alexander}\
  \bibnamefont {Morozov}}\ and\ \bibinfo {author} {\bibfnamefont {Davide}\
  \bibnamefont {Marenduzzo}},\ }\bibfield  {title} {\enquote {\bibinfo {title}
  {Enhanced diffusion of tracer particles in dilute bacterial suspensions},}\
  }\href@noop {} {\bibfield  {journal} {\bibinfo  {journal} {Soft Matter}\
  }\textbf {\bibinfo {volume} {10}},\ \bibinfo {pages} {2748--2758} (\bibinfo
  {year} {2014})}\BibitemShut {NoStop}%
\bibitem [{\citenamefont {Saintillan}\ and\ \citenamefont
  {Shelley}(2007)}]{Shelley2007}%
  \BibitemOpen
  \bibfield  {author} {\bibinfo {author} {\bibfnamefont {D.}~\bibnamefont
  {Saintillan}}\ and\ \bibinfo {author} {\bibfnamefont {M.~J.}\ \bibnamefont
  {Shelley}},\ }\bibfield  {title} {\enquote {\bibinfo {title} {Orientational
  order and instabilities in suspensions of self-locomoting rods},}\
  }\href@noop {} {\bibfield  {journal} {\bibinfo  {journal} {Physical Review
  Letters}\ }\textbf {\bibinfo {volume} {99}},\ \bibinfo {pages} {058102}
  (\bibinfo {year} {2007})}\BibitemShut {NoStop}%
\bibitem [{\citenamefont {Underhill}\ \emph {et~al.}(2008)\citenamefont
  {Underhill}, \citenamefont {Hernandez-Ortiz},\ and\ \citenamefont
  {Graham}}]{Underhill2008}%
  \BibitemOpen
  \bibfield  {author} {\bibinfo {author} {\bibfnamefont {P.~T.}\ \bibnamefont
  {Underhill}}, \bibinfo {author} {\bibfnamefont {J.~P.}\ \bibnamefont
  {Hernandez-Ortiz}}, \ and\ \bibinfo {author} {\bibfnamefont {M.~D.}\
  \bibnamefont {Graham}},\ }\bibfield  {title} {\enquote {\bibinfo {title}
  {Diffusion and spatial correlations in suspensions of swimming particles},}\
  }\href@noop {} {\bibfield  {journal} {\bibinfo  {journal} {Physical review
  letters}\ }\textbf {\bibinfo {volume} {100}},\ \bibinfo {pages} {248101}
  (\bibinfo {year} {2008})}\BibitemShut {NoStop}%
\bibitem [{\citenamefont {Lauga}\ and\ \citenamefont
  {Goldstein}(2012)}]{lauga_dance_2012}%
  \BibitemOpen
  \bibfield  {author} {\bibinfo {author} {\bibfnamefont {Eric}\ \bibnamefont
  {Lauga}}\ and\ \bibinfo {author} {\bibfnamefont {Raymond~E.}\ \bibnamefont
  {Goldstein}},\ }\bibfield  {title} {\enquote {\bibinfo {title} {Dance of the
  microswimmers. ({Cover} story)},}\ }\href@noop {} {\bibfield  {journal}
  {\bibinfo  {journal} {Physics Today}\ }\textbf {\bibinfo {volume} {65}},\
  \bibinfo {pages} {30--35} (\bibinfo {year} {2012})}\BibitemShut {NoStop}%
\bibitem [{\citenamefont {Gyrya}\ \emph {et~al.}(2010)\citenamefont {Gyrya},
  \citenamefont {Aranson}, \citenamefont {Berlyand},\ and\ \citenamefont
  {Karpeev}}]{gyrya_model_2010}%
  \BibitemOpen
  \bibfield  {author} {\bibinfo {author} {\bibfnamefont {Vitaliy}\ \bibnamefont
  {Gyrya}}, \bibinfo {author} {\bibfnamefont {Igor}\ \bibnamefont {Aranson}},
  \bibinfo {author} {\bibfnamefont {Leonid}\ \bibnamefont {Berlyand}}, \ and\
  \bibinfo {author} {\bibfnamefont {Dmitry}\ \bibnamefont {Karpeev}},\
  }\bibfield  {title} {\enquote {\bibinfo {title} {A {Model} of {Hydrodynamic}
  {Interaction} {Between} {Swimming} {Bacteria}},}\ }\href@noop {} {\bibfield
  {journal} {\bibinfo  {journal} {Bulletin of Mathematical Biology}\ }\textbf
  {\bibinfo {volume} {72}},\ \bibinfo {pages} {148--183} (\bibinfo {year}
  {2010})}\BibitemShut {NoStop}%
\bibitem [{\citenamefont {Michelin}\ and\ \citenamefont
  {Lauga}(2009)}]{Michelin2CoupledSwimmer2009}%
  \BibitemOpen
  \bibfield  {author} {\bibinfo {author} {\bibfnamefont {S{\'e}bastien}\
  \bibnamefont {Michelin}}\ and\ \bibinfo {author} {\bibfnamefont {Eric}\
  \bibnamefont {Lauga}},\ }\bibfield  {title} {\enquote {\bibinfo {title} {The
  long-time dynamics of two hydrodynamically-coupled swimming cells},}\
  }\href@noop {} {\bibfield  {journal} {\bibinfo  {journal} {Bulletin of
  Mathematical Biology}\ }\textbf {\bibinfo {volume} {72}},\ \bibinfo {pages}
  {973--1005} (\bibinfo {year} {2009})}\BibitemShut {NoStop}%
\bibitem [{\citenamefont {Thiffeault}(2015)}]{ThiffeaultNonGuassian2015}%
  \BibitemOpen
  \bibfield  {author} {\bibinfo {author} {\bibfnamefont {Jean-Luc}\
  \bibnamefont {Thiffeault}},\ }\bibfield  {title} {\enquote {\bibinfo {title}
  {Distribution of particle displacements due to swimming microorganisms},}\
  }\href@noop {} {\bibfield  {journal} {\bibinfo  {journal} {Phys. Rev. E}\
  }\textbf {\bibinfo {volume} {92}},\ \bibinfo {pages} {023023} (\bibinfo
  {year} {2015})}\BibitemShut {NoStop}%
\bibitem [{\citenamefont {Thiffeault}\ and\ \citenamefont
  {Childress}(2010)}]{ThiffeaultSquirmers2010}%
  \BibitemOpen
  \bibfield  {author} {\bibinfo {author} {\bibfnamefont {Jean-Luc}\
  \bibnamefont {Thiffeault}}\ and\ \bibinfo {author} {\bibfnamefont {Stephen}\
  \bibnamefont {Childress}},\ }\bibfield  {title} {\enquote {\bibinfo {title}
  {Stirring by swimming bodies},}\ }\href@noop {} {\bibfield  {journal}
  {\bibinfo  {journal} {Physics Letters A}\ }\textbf {\bibinfo {volume}
  {374}},\ \bibinfo {pages} {3487 -- 3490} (\bibinfo {year}
  {2010})}\BibitemShut {NoStop}%
\bibitem [{\citenamefont {Stenhammar}\ \emph {et~al.}(2014)\citenamefont
  {Stenhammar}, \citenamefont {Marenduzzo}, \citenamefont {Allen},\ and\
  \citenamefont {Cates}}]{CatesABPDimensionality2014}%
  \BibitemOpen
  \bibfield  {author} {\bibinfo {author} {\bibfnamefont {Joakim}\ \bibnamefont
  {Stenhammar}}, \bibinfo {author} {\bibfnamefont {Davide}\ \bibnamefont
  {Marenduzzo}}, \bibinfo {author} {\bibfnamefont {Rosalind~J.}\ \bibnamefont
  {Allen}}, \ and\ \bibinfo {author} {\bibfnamefont {Michael~E.}\ \bibnamefont
  {Cates}},\ }\bibfield  {title} {\enquote {\bibinfo {title} {Phase behaviour
  of active {Brownian} particles: the role of dimensionality},}\ }\href@noop {}
  {\bibfield  {journal} {\bibinfo  {journal} {Soft Matter}\ }\textbf {\bibinfo
  {volume} {10}},\ \bibinfo {pages} {1489--1499} (\bibinfo {year}
  {2014})}\BibitemShut {NoStop}%
\bibitem [{\citenamefont {Dean}(1996)}]{Dean1996}%
  \BibitemOpen
  \bibfield  {author} {\bibinfo {author} {\bibfnamefont {David~S}\ \bibnamefont
  {Dean}},\ }\bibfield  {title} {\enquote {\bibinfo {title} {Langevin equation
  for the density of a system of interacting langevin processes},}\ }\href@noop
  {} {\bibfield  {journal} {\bibinfo  {journal} {J. Phys. A: Math. Gen.}\
  }\textbf {\bibinfo {volume} {29}},\ \bibinfo {pages} {L613--L617} (\bibinfo
  {year} {1996})}\BibitemShut {NoStop}%
\bibitem [{\citenamefont {Das}\ and\ \citenamefont
  {Yoshimori}(2013)}]{Das2013}%
  \BibitemOpen
  \bibfield  {author} {\bibinfo {author} {\bibfnamefont {Shankar~P.}\
  \bibnamefont {Das}}\ and\ \bibinfo {author} {\bibfnamefont {Akira}\
  \bibnamefont {Yoshimori}},\ }\bibfield  {title} {\enquote {\bibinfo {title}
  {Coarse-grained forms for equations describing the microscopic motion of
  particles in a fluid},}\ }\href@noop {} {\bibfield  {journal} {\bibinfo
  {journal} {Phys. Rev. E}\ }\textbf {\bibinfo {volume} {88}},\ \bibinfo
  {pages} {043008} (\bibinfo {year} {2013})}\BibitemShut {NoStop}%
\bibitem [{\citenamefont {Lef{\`e}vre}\ and\ \citenamefont
  {Biroli}(2007)}]{JstatmechStoField2007}%
  \BibitemOpen
  \bibfield  {author} {\bibinfo {author} {\bibfnamefont {Alexandre}\
  \bibnamefont {Lef{\`e}vre}}\ and\ \bibinfo {author} {\bibfnamefont {Giulio}\
  \bibnamefont {Biroli}},\ }\bibfield  {title} {\enquote {\bibinfo {title}
  {Dynamics of interacting particle systems: stochastic process and field
  theory},}\ }\href@noop {} {\bibfield  {journal} {\bibinfo  {journal} {Journal
  of Statistical Mechanics: Theory and Experiment}\ }\textbf {\bibinfo {volume}
  {2007}},\ \bibinfo {pages} {P07024} (\bibinfo {year} {2007})}\BibitemShut
  {NoStop}%
\bibitem [{\citenamefont {Tailleur}\ and\ \citenamefont
  {Cates}(2008)}]{CatePRL2008}%
  \BibitemOpen
  \bibfield  {author} {\bibinfo {author} {\bibfnamefont {J.}~\bibnamefont
  {Tailleur}}\ and\ \bibinfo {author} {\bibfnamefont {M.~E.}\ \bibnamefont
  {Cates}},\ }\bibfield  {title} {\enquote {\bibinfo {title} {Statistical
  mechanics of interacting run-and-tumble bacteria},}\ }\href@noop {}
  {\bibfield  {journal} {\bibinfo  {journal} {Phys. Rev. Lett.}\ }\textbf
  {\bibinfo {volume} {100}},\ \bibinfo {pages} {218103} (\bibinfo {year}
  {2008})}\BibitemShut {NoStop}%
\bibitem [{\citenamefont {Solon}\ \emph {et~al.}(2015)\citenamefont {Solon},
  \citenamefont {Cates},\ and\ \citenamefont {Tailleur}}]{Solon2015}%
  \BibitemOpen
  \bibfield  {author} {\bibinfo {author} {\bibfnamefont {A.~P.}\ \bibnamefont
  {Solon}}, \bibinfo {author} {\bibfnamefont {M.~E.}\ \bibnamefont {Cates}}, \
  and\ \bibinfo {author} {\bibfnamefont {J.}~\bibnamefont {Tailleur}},\
  }\bibfield  {title} {\enquote {\bibinfo {title} {Active brownian particles
  and run-and-tumble particles: A comparative study},}\ }\href@noop {}
  {\bibfield  {journal} {\bibinfo  {journal} {The European Physical Journal
  Special Topics}\ }\textbf {\bibinfo {volume} {224}},\ \bibinfo {pages}
  {1231--1262} (\bibinfo {year} {2015})}\BibitemShut {NoStop}%
\bibitem [{\citenamefont {Lau}\ and\ \citenamefont
  {Lubensky}(2009)}]{Lubensky2009}%
  \BibitemOpen
  \bibfield  {author} {\bibinfo {author} {\bibfnamefont {A.~W.~C.}\
  \bibnamefont {Lau}}\ and\ \bibinfo {author} {\bibfnamefont {T.~C.}\
  \bibnamefont {Lubensky}},\ }\bibfield  {title} {\enquote {\bibinfo {title}
  {Fluctuating hydrodynamics and microrheology of a dilute suspension of
  swimming bacteria},}\ }\href@noop {} {\bibfield  {journal} {\bibinfo
  {journal} {Phys. Rev. E}\ }\textbf {\bibinfo {volume} {80}},\ \bibinfo
  {pages} {011917} (\bibinfo {year} {2009})}\BibitemShut {NoStop}%
\bibitem [{\citenamefont {Chavanis}(2008)}]{Chavanis2008}%
  \BibitemOpen
  \bibfield  {author} {\bibinfo {author} {\bibfnamefont {Pierre-Henri}\
  \bibnamefont {Chavanis}},\ }\bibfield  {title} {\enquote {\bibinfo {title}
  {Hamiltonian and {Brownian} systems with long-range interactions: {V}.
  {Stochastic} kinetic equations and theory of fluctuations},}\ }\href@noop {}
  {\bibfield  {journal} {\bibinfo  {journal} {Physica A: Statistical Mechanics
  and its Applications}\ }\textbf {\bibinfo {volume} {387}},\ \bibinfo {pages}
  {5716--5740} (\bibinfo {year} {2008})}\BibitemShut {NoStop}%
\bibitem [{\citenamefont {Lau}\ and\ \citenamefont
  {Lubensky}(2007)}]{Lubensky20072}%
  \BibitemOpen
  \bibfield  {author} {\bibinfo {author} {\bibfnamefont {A.~W.~C.}\
  \bibnamefont {Lau}}\ and\ \bibinfo {author} {\bibfnamefont {T.~C.}\
  \bibnamefont {Lubensky}},\ }\bibfield  {title} {\enquote {\bibinfo {title}
  {State-dependent diffusion: Thermodynamic consistency and its path integral
  formulation},}\ }\href@noop {} {\bibfield  {journal} {\bibinfo  {journal}
  {Phys. Rev. E}\ }\textbf {\bibinfo {volume} {76}},\ \bibinfo {pages} {011123}
  (\bibinfo {year} {2007})}\BibitemShut {NoStop}%
\bibitem [{\citenamefont {Bozorgi}\ and\ \citenamefont
  {Underhill}(2011)}]{Bozorgi2011}%
  \BibitemOpen
  \bibfield  {author} {\bibinfo {author} {\bibfnamefont {Yaser}\ \bibnamefont
  {Bozorgi}}\ and\ \bibinfo {author} {\bibfnamefont {Patrick~T.}\ \bibnamefont
  {Underhill}},\ }\bibfield  {title} {\enquote {\bibinfo {title} {Effect of
  viscoelasticity on the collective behavior of swimming microorganisms},}\
  }\href@noop {} {\bibfield  {journal} {\bibinfo  {journal} {Phys. Rev. E}\
  }\textbf {\bibinfo {volume} {84}},\ \bibinfo {pages} {061901} (\bibinfo
  {year} {2011})}\BibitemShut {NoStop}%
\bibitem [{\citenamefont {Heidenreich}\ \emph {et~al.}(2011)\citenamefont
  {Heidenreich}, \citenamefont {Hess},\ and\ \citenamefont
  {Klapp}}]{KlappNonlinear2011}%
  \BibitemOpen
  \bibfield  {author} {\bibinfo {author} {\bibfnamefont {Sebastian}\
  \bibnamefont {Heidenreich}}, \bibinfo {author} {\bibfnamefont {Siegfried}\
  \bibnamefont {Hess}}, \ and\ \bibinfo {author} {\bibfnamefont {Sabine H.~L.}\
  \bibnamefont {Klapp}},\ }\bibfield  {title} {\enquote {\bibinfo {title}
  {Nonlinear rheology of active particle suspensions: Insights from an
  analytical approach},}\ }\href@noop {} {\bibfield  {journal} {\bibinfo
  {journal} {Phys Rev E}\ }\textbf {\bibinfo {volume} {83}},\ \bibinfo {pages}
  {011907} (\bibinfo {year} {2011})}\BibitemShut {NoStop}%
\bibitem [{\citenamefont {Baskaran}\ and\ \citenamefont
  {Cristina~Marchetti}(2010)}]{Marchetti_2010}%
  \BibitemOpen
  \bibfield  {author} {\bibinfo {author} {\bibfnamefont {Aparna}\ \bibnamefont
  {Baskaran}}\ and\ \bibinfo {author} {\bibfnamefont {M}~\bibnamefont
  {Cristina~Marchetti}},\ }\bibfield  {title} {\enquote {\bibinfo {title}
  {Nonequilibrium statistical mechanics of self-propelled hard rods},}\
  }\href@noop {} {\bibfield  {journal} {\bibinfo  {journal} {Journal of
  Statistical Mechanics: Theory and Experiment}\ }\textbf {\bibinfo {volume}
  {2010}},\ \bibinfo {pages} {P04019} (\bibinfo {year} {2010})}\BibitemShut
  {NoStop}%
\bibitem [{\citenamefont {Furukawa}\ \emph {et~al.}(2014)\citenamefont
  {Furukawa}, \citenamefont {Marenduzzo},\ and\ \citenamefont
  {Cates}}]{CatesCluster2014}%
  \BibitemOpen
  \bibfield  {author} {\bibinfo {author} {\bibfnamefont {Akira}\ \bibnamefont
  {Furukawa}}, \bibinfo {author} {\bibfnamefont {Davide}\ \bibnamefont
  {Marenduzzo}}, \ and\ \bibinfo {author} {\bibfnamefont {Michael~E.}\
  \bibnamefont {Cates}},\ }\bibfield  {title} {\enquote {\bibinfo {title}
  {Activity-induced clustering in model dumbbell swimmers: {The} role of
  hydrodynamic interactions},}\ }\href@noop {} {\bibfield  {journal} {\bibinfo
  {journal} {Physical Review E}\ }\textbf {\bibinfo {volume} {90}},\ \bibinfo
  {pages} {022303} (\bibinfo {year} {2014})}\BibitemShut {NoStop}%
\bibitem [{\citenamefont {Kubo}(1963)}]{rk:sle}%
  \BibitemOpen
  \bibfield  {author} {\bibinfo {author} {\bibfnamefont {Ryogo}\ \bibnamefont
  {Kubo}},\ }\bibfield  {title} {\enquote {\bibinfo {title} {Stochastic
  {L}iouville equations},}\ }\href@noop {} {\bibfield  {journal} {\bibinfo
  {journal} {J. Mathematical Phys.}\ }\textbf {\bibinfo {volume} {4}},\
  \bibinfo {pages} {174--183} (\bibinfo {year} {1963})}\BibitemShut {NoStop}%
\bibitem [{\citenamefont {Leptos}\ \emph {et~al.}(2009)\citenamefont {Leptos},
  \citenamefont {Guasto}, \citenamefont {Gollub}, \citenamefont {Pesci},\ and\
  \citenamefont {Goldstein}}]{GoldsteinDiffusion2009}%
  \BibitemOpen
  \bibfield  {author} {\bibinfo {author} {\bibfnamefont {Kyriacos~C.}\
  \bibnamefont {Leptos}}, \bibinfo {author} {\bibfnamefont {Jeffrey~S.}\
  \bibnamefont {Guasto}}, \bibinfo {author} {\bibfnamefont {J.~P.}\
  \bibnamefont {Gollub}}, \bibinfo {author} {\bibfnamefont {Adriana~I.}\
  \bibnamefont {Pesci}}, \ and\ \bibinfo {author} {\bibfnamefont {Raymond~E.}\
  \bibnamefont {Goldstein}},\ }\bibfield  {title} {\enquote {\bibinfo {title}
  {Dynamics of enhanced tracer diffusion in suspensions of swimming eukaryotic
  microorganisms},}\ }\href@noop {} {\bibfield  {journal} {\bibinfo  {journal}
  {Phys. Rev. Lett.}\ }\textbf {\bibinfo {volume} {103}},\ \bibinfo {pages}
  {198103} (\bibinfo {year} {2009})}\BibitemShut {NoStop}%
\bibitem [{\citenamefont {Caspi}\ \emph {et~al.}(2000)\citenamefont {Caspi},
  \citenamefont {Granek},\ and\ \citenamefont {Elbaum}}]{caspi2000enhanced}%
  \BibitemOpen
  \bibfield  {author} {\bibinfo {author} {\bibfnamefont {Avi}\ \bibnamefont
  {Caspi}}, \bibinfo {author} {\bibfnamefont {Rony}\ \bibnamefont {Granek}}, \
  and\ \bibinfo {author} {\bibfnamefont {Michael}\ \bibnamefont {Elbaum}},\
  }\bibfield  {title} {\enquote {\bibinfo {title} {Enhanced diffusion in active
  intracellular transport},}\ }\href@noop {} {\bibfield  {journal} {\bibinfo
  {journal} {Physical Review Letters}\ }\textbf {\bibinfo {volume} {85}},\
  \bibinfo {pages} {5655} (\bibinfo {year} {2000})}\BibitemShut {NoStop}%
\end{thebibliography}

%

\end{document}